\documentclass[10pt,conference]{IEEEtran}
\IEEEoverridecommandlockouts

\usepackage[left=0.65in,right=0.65in,top=1in,bottom=1in]{geometry}

\usepackage[colorlinks=true,allcolors=blue,bookmarks=false]{hyperref}
\usepackage{array}
\usepackage{tabularx}
\usepackage{multirow}
\newcolumntype{L}[1]{>{\raggedright\arraybackslash}m{#1}}
\newcolumntype{C}[1]{>{\centering\arraybackslash}m{#1}}
\newcolumntype{R}[1]{>{\raggedleft\arraybackslash}m{#1}}
\usepackage{svg}
\usepackage{balance}
\usepackage{cite}
\usepackage{amsmath,amssymb,amsfonts}
\usepackage{pgfplots}
\usepackage{url}
\pgfplotsset{compat=newest}
\usepackage{algorithm,algpseudocode}
\usepackage{graphicx}
\usepackage{textcomp}
\usepackage{adjustbox}
\usepackage{xcolor}
\usepackage{tikz}
\usepackage{tabularx} 
\usepackage{booktabs} 
\usepackage{float}

\def\BibTeX{{\rm B\kern-.05em{\sc i\kern-.025em b}\kern-.08em
    T\kern-.1667em\lower.7ex\hbox{E}\kern-.125emX}}

\begin{document}

\title{REDUS: Adaptive Resampling for Efficient Deep Learning in Centralized and Federated IoT Networks
\thanks{This work was
supported in part by the Natural Sciences and Engineering Research Council of Canada under Award RGPIN-2020-06260; and in part by the National Science Foundation of the USA under Awards 2210251 and 2210252.}
}

\author{
	\IEEEauthorblockN{
        Eyad Gad\IEEEauthorrefmark{1}$^1$, 
	Gad Gad\IEEEauthorrefmark{1}$^2$, 
    Mostafa M. Fouda\IEEEauthorrefmark{2}\IEEEauthorrefmark{3}$^3$, Mohamed I. Ibrahem\IEEEauthorrefmark{4}$^4$, Muhammad Ismail\IEEEauthorrefmark{5}$^5$, and \\Zubair Md Fadlullah\IEEEauthorrefmark{1}$^6$
 }

\IEEEauthorblockA{\IEEEauthorrefmark{1}Department of Computer Science, Western University, London, ON, Canada.}
\IEEEauthorblockA{\IEEEauthorrefmark{2}Department of Electrical and Computer Engineering, Idaho State University, Pocatello, ID, USA.}
\IEEEauthorblockA{\IEEEauthorrefmark{3}Center for Advanced Energy Studies (CAES), Idaho Falls, ID, USA.}
\IEEEauthorblockA{\IEEEauthorrefmark{4}School of Computer and Cyber Sciences, Augusta University, Augusta, GA, USA.}
\IEEEauthorblockA{\IEEEauthorrefmark{5}Department of Computer Science, Tennessee Technological University, Cookeville, TN, USA.}

Emails:
$^1$egad@uwo.ca,
$^2$ggad@uwo.ca,
$^3$mfouda@ieee.org,
$^4$mibrahem@augusta.edu,
$^5$mismail@tntech.edu,\\
$^6$zfadlullah@ieee.org
} 

\maketitle

\begin{abstract}
With the rise of Software-Defined Networking (SDN) for managing traffic and ensuring seamless operations across interconnected devices, challenges arise when SDN controllers share infrastructure with deep learning (DL) workloads. Resource contention between DL training and SDN operations, especially in latency-sensitive IoT environments, can degrade SDN's responsiveness and compromise network performance. Federated Learning (FL) helps address some of these concerns by decentralizing DL training to edge devices, thus reducing data transmission costs and enhancing privacy. Yet, the computational demands of DL training can still interfere with SDN's performance, especially under the continuous data streams characteristic of IoT systems. To mitigate this issue, we propose REDUS (Resampling for Efficient Data Utilization in Smart-Networks), a resampling technique that optimizes DL training by prioritizing misclassified samples and excluding redundant data, inspired by AdaBoost. REDUS reduces the number of training samples per epoch, thereby conserving computational resources, reducing energy consumption, and accelerating convergence without significantly impacting accuracy. Applied within an FL setup, REDUS enhances the efficiency of model training on resource-limited edge devices while maintaining network performance. In this paper, REDUS is evaluated on the CICIoT2023 dataset for IoT attack detection, showing a training time reduction of up to 72.6\% with a minimal accuracy loss of only 1.62\%, offering a scalable and practical solution for intelligent networks.

\end{abstract}


\section{Introduction}
Modern networks increasingly depend on Software-Defined Networking (SDN) to dynamically manage traffic, enhance real-time performance, and ensure smooth operations across interconnected devices~\cite{9717267}. While SDN controllers deployed on shared servers offer flexible network control, they often coexist with deep learning (DL) workloads running on the same infrastructure. This simultaneous execution of resource-intensive tasks introduces resource contention, as DL training demands significant CPU, memory, and bandwidth. Such resource competition can disrupt latency-sensitive SDN operations, impairing real-time traffic management, security enforcement, and policy execution.

The rising demand for data-driven intelligent systems has fueled the use of DL models in IoT (Internet of Things) attack detection. These models are traditionally trained on centralized servers, which poses several challenges, including increased communication overhead, latency, and privacy risks from centralizing sensitive data~\cite{10736556}. Federated Learning (FL) offers a solution by reducing communication loads, improving privacy, and better utilizing edge resources~\cite{zhang2021survey,9060924, 9764093,10000881, 9500351, 9647636}. However, DL training’s computational demands can still interfere with SDN's computational capacity, especially during peak loads. This is further complicated in IoT settings, where continuous data streams from edge devices strain both network management and resource allocation~\cite{9975908,NASSER2022108672}.

To address these challenges, we propose REDUS, a novel resampling technique aimed at optimizing DL workloads. REDUS minimizes redundancy in DL training by gradually reducing the number of training samples per epoch, focusing the training load on misclassified samples that require further learning. By eliminating the repeated processing of already learned data, REDUS reduces the training time, thus reducing computational costs, and accelerates convergence. It can be utilized to ensure the reliability of IoT networks, offering a practical solution for future intelligent environments.

To this end, we introduce REDUS, an AdaBoost-inspired resampling method designed for efficient training. Although REDUS draws inspiration from AdaBoost~\cite{Wang2019, 9712386}, it offers two key innovations that set it apart:

\begin{enumerate}
    \item In AdaBoost, sample weights are adjusted to train a new weak learner on the same dataset with modified weights. In contrast, REDUS updates sample weights at each epoch to train a single strong learner (i.e., a DL model) using fewer samples per epoch. Randomly excluding samples can reduce accuracy compared to conventional (vanilla) DL training, where in each epoch, the model performs backward propagation on each sample in the training dataset. However, REDUS reduces training time and thus power consumption on resource-constrained devices by selecting a focused subset of data with minimal impact on accuracy.

    \item Unlike AdaBoost, which builds an ensemble of weak learners whose combined votes form a strong model, REDUS applies to DL models trained using the iterative minimization of a loss function via Stochastic Gradient Descent (SGD). This approach is effective in both centralized and decentralized learning settings, particularly in environments where energy resources are limited, such as edge devices. REDUS ensures computational efficiency without sacrificing model performance, making it an ideal fit for FL systems.
\end{enumerate}

In this study, we evaluate the REDUS resampling approach within an FL setup, focusing on IoT attack detection using the CICIoT2023 dataset~\cite{ciciot23}. This dataset offers a collection of large-scale IoT attacks, designed to provide a realistic testbed for assessing the performance of intrusion detection systems in IoT environments.

The rest of the paper is organized as follows. In Section \ref{sec:background}, we provide a comprehensive review of related work, surveying existing methods and approaches relevant to our proposed REDUS resampling method. Section \ref{sec:prelim} offers background information by discussing AdaBoost, an ML algorithm that serves as the inspiration for our resampling technique. Our proposed REDUS resampling method is detailed in Section \ref{sec:methods} which is divided into two subsections,  Subsection \ref{sec:REDUS_method} introduces the REDUS method, while Subsection \ref{sec:REDUS_fed} extends its application to the FL setting. Section \ref{sec:exp_res} outlines our experimental setup and presents the empirical results. Finally, Section \ref{sec:conclusion} summarizes our findings and discusses the implications of the REDUS method.

\section{Related Work}
\label{sec:background}
In this section, we explore studies focused on minimizing training time, a critical factor for the efficiency and scalability of DL models. Tang \textit{et al.} \cite{tang2021data} investigated Non-Independent and Identically Distributed (Non-IID) label distributions within FL. Conventional imbalanced learning methods, which typically employ data resampling to increase sampling weights for minority classes, may not integrate smoothly with FL, potentially affecting convergence and accuracy. Tang \textit{et al.}'s analysis revealed two key findings: data resampling in FL can enhance convergence by balancing label sampling across clients, but it may also reduce local dataset accuracy due to overfitting or underfitting. To address these issues, they proposed Imbalanced Weight Decay Sampling (IWDS), a data resampling strategy that dynamically adjusts label sampling to optimize convergence while preserving local data knowledge. IWDS improves FL algorithms without additional computational complexities or requiring inter-client communication.

Additionally, Taherkhani \textit{et al.} \cite{TAHERKHANI2020351} introduced AdaBoost-CNN, a novel approach combining AdaBoost and Convolutional Neural Networks (CNN) for tackling imbalanced datasets. This framework leverages ensemble techniques, using AdaBoost’s sequential training of classifiers to improve accuracy via weighted samples. A distinctive aspect of this method is its use of transfer learning, allowing knowledge transfer between successive CNN estimators, thereby reducing computational load. Empirical results demonstrated AdaBoost-CNN’s superior accuracy on both synthetic imbalanced datasets and benchmarks like CIFAR-10 and Fashion-MNIST, with faster training times, offering a robust solution for imbalanced data challenges.

Lastly, Huang \textit{et al.} \cite{huang2020loAdaBoost} addressed intensive care data distribution in FL. Previous FL studies have mainly focused on test accuracy, privacy \cite{el2022differential, 9069945, 10.5555/3361338.3361469}, or communication efficiency \cite{almanifi2023communication, oh2022communication}, often overlooking computational load on client devices. To address this, the authors introduced Loss-based Adaptive Boosting Federated Averaging (LoAdaBoost FedAvg), optimizing local computational complexity, communication costs, and test accuracy. Evaluations using MIMIC-III and eICU databases showed LoAdaBoost FedAvg’s superior predictive accuracy and reduced computational demands, introducing an efficient algorithmic solution for critical health data in FL.

\section{Preliminaries}
\label{sec:prelim} 
In this section, we review AdaBoost, A Machine Learning (ML) algorithm that trains an ensemble of weak learners to form a strong learner. AdaBoost utilizes a sample weighting scheme that we inspire by our REDUS method.

The fundamental concept behind AdaBoost is to combine multiple ``weak learners" (simple models) to create a ``strong learner" (a highly accurate model).  Let \(H= \hat{h_f}\) be the set of constructed weak learners. Let's suppose training sample data is \({(x_1,y_1),......,(X_i,y_i),........,(x_n,y_n)}\) where \( x_i\) is the $i^{th}$ feature of the vector \(y_i\in {(+1,-1)}\). Let (\(w_i,w_2,.................w_n\))  be the sample weights that reflect the importance degrees of the samples and, in statistical terms, represent an estimation of the sample distribution.

The working of the AdaBoost algorithm is as follows:
\begin{itemize}
    \item  \textbf{ Step 1:}  Initialize weights \(w_i(1)\) ( \(i= 1,.............., n\)) satisfying \(\sum_{i=1}^{n} w_i(1)=1\)
     Initially, AdaBoost assigns equal weights to all instances in the training set. So, the initial weight of each instance 
    \(w_i= 1/n\), where \(n\) is the total number of instances.

    \item \textbf{ Step 2:} For each iteration \(t\) wether \(t\) = \({1,2,............,T}\)  and \(T\) is total number of iterations. A weak learner (e.g., a decision stump) is trained on the weighted training data.
     The error \(\epsilon_t\) of an \(t\) iteration of the weak learner is calculated as:
     \begin{equation}
         \epsilon_t= \frac{\sum_{i=1}^{n}.w_i.I(y_i\neq h_t(x_i))}{\sum{i=1}^{n} w_i} ,
     \end{equation}
     where \(I\) is a function that returns \(1\), if the sample \(x_i\) is misclassified and \(0\) otherwise; \(h_t(x_i)\) is the prediction of the weak learner.
     \item {Step 3:} The weights of the instances are updated based on the performance of the weak learner. If an instance is correctly classified, its weight is decreased, and if it is misclassified, its weight is increased. The weight update rule is:
     \begin{equation}
         w_i \leftarrow w_i \cdot \exp(\alpha_t \cdot I(y_i \neq h_t(x_i))),
     \end{equation}
      where \(\alpha_t =\frac{1}{2} \log(\frac{1-\epsilon_t}{\epsilon_t})\) is the weight of the weak learner in the final model.

    \item \textbf{Step 4:} After updating the weights, they are normalized so that their sum is \(1\).
    \item  \textbf{ Step 5:}  The final model is a weighted combination of the weak learners:
    \begin{equation}
        H_(x)= sign(\sum_{t=1}^{T} \alpha_t.h_t(x)).
    \end{equation}
    
\end{itemize}

\begin{figure}[!t]
    \centering
    \includesvg[width=0.8\linewidth]{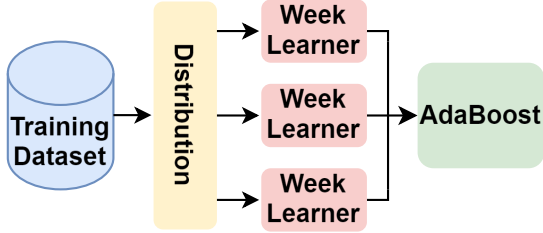}
    \caption{Overview of the Adaboost algorithm.}
    \label{fig:AdaBoost}
\end{figure}

It is important to note that while AdaBoost can involve a process that resembles resampling, its primary goal is to improve classification performance by focusing on the more complex cases in the dataset. The 'resampling' in AdaBoost is more about reweighting and focusing on challenging instances rather than creating independent data samples. Fig.~\ref{fig:AdaBoost} shows the working of AdaBoost algorithm.

\section{Proposed Method}
\label{sec:methods}
\begin{figure}[!t]
    \centering
    \includegraphics[width=\linewidth]{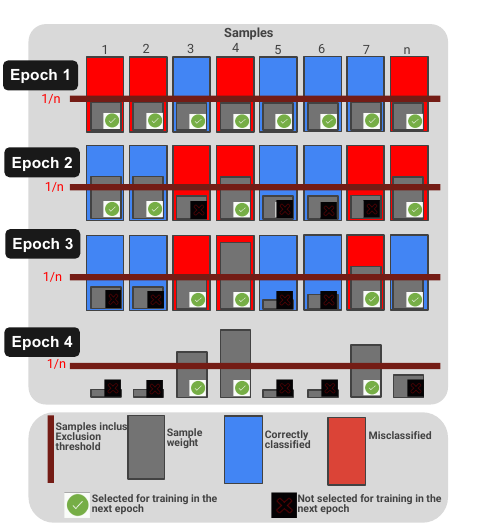}
     \vspace{-0.3cm}
    \caption{An overview of the REDUS training. The figure shows training a model on n samples for four epochs. Each sample has an initial weight of $1/n$. As training progresses, samples' weights change based on classification results. When a sample's assigned weight drops below a pre-fixed threshold, the sample is excluded from training in the next epoch.}
    \label{fig-redus}
\end{figure}

In this section, we first propose our resampling method dubbed REDUS and then present how it is integrated in a FL system. The workflow of REDUS is depicted in Fig.~\ref{fig-redus}. 
\subsection{The Proposed Resampling Method}
\label{sec:REDUS_method}


In this section, we describe our proposed resampling approach. Recall that in vanilla DL training, the goal is to train a model $f$ on a prediction task using a dataset $D$ which is composed of $n$ (input, label) pairs: ($x_i$, $y_i$).  The label $y^i \in \{1, 2, \ldots, C\}$ is a one-hot encoded vector of size $C$, C, where $C$ is the number of classes in $D$. DL training involves iteratively updating the weights $\mathbf{w}$ of the model $f$ such that, given an input \(x_i\), the predicted output \(\hat{y}_i = f(x_i)\) approximates the true label \(y_i\). 

This is achieved by iteratively calculating the loss between \(\hat{y}_i\) and \(y_i\), and computing the gradient of the loss. The loss calculated over the dataset is defined as:

\begin{equation}
\mathcal{L} = \frac{1}{n} \sum_{i=1}^{n} L_{CE}(f; x_i, y_i),
\label{eq:loss}
\end{equation}
where \(L_{CE}\) denotes the cross-entropy loss function, defined as:
\[
L_{CE}(f; x, y) = - \sum_{c=1}^{C} y_{c} \log(f_{c}(x)),
\]
 \(f_{c}(x)\) is the $c^{th}$ element of the output probability vector produced by the model \(f\) for input \(x\), and \(y\) is the corresponding one-hot encoded label.
\begin{algorithm}
\caption{DL Training using REDUS Method}
\label{alg:REDUS}
\begin{algorithmic}[1]
    \For{$t = 1$ to $E$}
        \State $\hat{D} \gets []$
        \If{$t == 1$}
            \For{$i = 1$ to $|D|$}
                \State $w_i \gets \frac{1}{n}$
            \EndFor
        \Else
            \State $\varepsilon_t \gets \sum_{i}^{n} w_i \cdot [1_{f(x_i) \neq y_i}]$
            \State $\alpha_t \gets \frac{1}{2} \log\left(\frac{1 - \varepsilon_t}{\varepsilon_t}\right)$
            \State $Z_t \gets 2 \sqrt{\varepsilon_t(1 - \varepsilon_t)}$
            \For{$i = 1$ to $|D|$}
                \State $w_i \gets
                    \begin{cases} 
                        \frac{w_i \cdot e^{\alpha_t}}{Z_t} & \text{if } f(x_i) \neq y_i \\
                        \frac{w_i \cdot e^{-\alpha_t}}{Z_t} & \text{if } f(x_i) = y_i
                    \end{cases}$
            \EndFor
        \EndIf
        \For{$i = 1$ to $|D|$}
            \If {$w_i \geq \theta$}
                \State $\hat{D}.append((x_i, y_i))$
            \EndIf 
        \EndFor

        \State Train model $f$ on $\hat{D}$ using equations \ref{eq:loss} and \ref{eq:w_update}
    \EndFor
\end{algorithmic}
\end{algorithm}

 The weights $\mathbf{w}$ are then updated using the gradient of the loss with respect to the weights:

\begin{equation}
    \mathbf{w} \leftarrow \mathbf{w} - \eta \nabla_{\mathbf{w}} \mathcal{L},
    \label{eq:w_update}
\end{equation}
where $\eta$ is the learning rate, and $\nabla_{\mathbf{w}} \mathcal{L}$ is the gradient of the loss $\mathcal{L}$ with respect to the weights $\mathbf{w}$.

In each epoch of vanilla DL training, forward and back-propagation operations are applied to the model $f$ on all samples in the dataset $D$. We denote the back-propagation time for one sample as $\tau$. Then, the total training time $(LLT)$ is given by:
\begin{equation}
    LTT = |D| \cdot \tau \cdot E ,
\end{equation} 
where $E$ is the number of epochs.
The goal of our proposed resampling approach is to gradually reduce the number of samples used in training during each epoch, thereby reducing the total training time as defined by $LTT$.
As illustrated in Algorithm~\ref{alg:REDUS}, before training starts, each sample in the training dataset \(D\) is assigned an initial weight \(w\) as:  
\begin{equation}
    w_i \gets \frac{1}{n}, \quad \forall i \in D.
\end{equation}

$w_i$ is updated at each epoch and controls whether the $i^{th}$ sample is included in training. If $w_i$ decreases below a certain threshold $\theta$, the sample will be excluded from training. However, even after the $i^{\text{th}}$ sample is excluded, its weight $w_i$ continues to be updated each epoch. If $w_i$ exceeds the threshold $\theta$ during a re-evaluation in any subsequent epoch, the sample will be re-included in the training set. The rules for updating $w_i$ are given by: 
\begin{equation}
w_i \gets
\begin{cases} 
    \frac{w_i \cdot e^{\alpha_t}}{Z_t} & \text{if } f(x_i) \neq y_i, \\
    \frac{w_i \cdot e^{-\alpha_t}}{Z_t} & \text{if } f(x_i) = y_i,
\end{cases}
\end{equation}

where $\alpha$ and $Z_t$ are given by:
\begin{equation}
    \varepsilon_t \gets \sum_{i}^{n} w_i \cdot [1_{f(x_i) \neq y_i}],
\end{equation}
\begin{equation}
    \alpha_t \gets \frac{1}{2} \log\left(\frac{1 - \varepsilon_t}{\varepsilon_t}\right),
\end{equation}
\begin{equation}
    Z_t \gets 2 \sqrt{\varepsilon_t(1 - \varepsilon_t)}.
\end{equation}

Instead of training on the training dataset $D$, a new dataset $\hat{D}$ is created at each epoch, consisting only of samples whose weights exceed a threshold $\theta$. The model $f$ is then trained for one epoch on $\hat{D}$. The time complexity of the REDUS method is \(O(E \cdot n)\).

\begin{figure}[!t]
    \centering
    \includesvg[width=\linewidth]{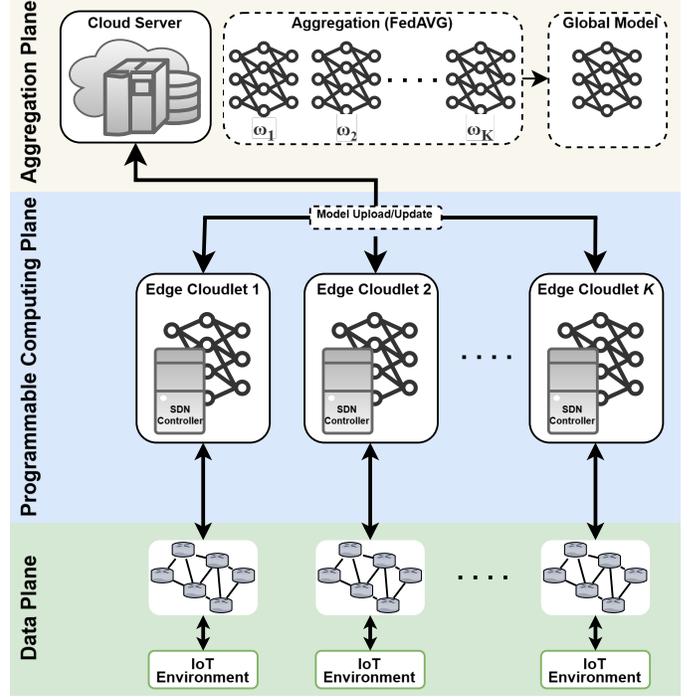} 
    \vspace{-0.5cm}
    \caption{Overview of the FL architecture adopted in this study.}
    
    \label{fig:fl_example}
    \vspace{-0.4cm}
\end{figure}

The threshold $\theta$ ranges from $0$ to $1/n$. Setting $\theta = 0$ includes all samples in training, as all sample weights are greater than or equal to zero. As $\theta$ increases, only samples with the highest weights are included in the training. During training, misclassified samples, which are harder to learn, receive higher weights, thereby prioritizing them in subsequent epochs. Conversely, correctly classified samples see their weights decrease. Once their weights fall below $\theta$, these samples are excluded from training, as illustrated in Fig.~\ref{fig:fl_example}.

\subsection{Integration of REDUS With FL}
\label{sec:REDUS_fed}

In this section, we extend the REDUS method to the FL setting. FL systems, by design, distribute the training of a global model across a set of $K$ clients, each of which $i$ trains the model locally on its own dataset $D_i$. At each communication round $r$, the client $i$ receives the global model $g_i^r$ from the central server. The client then trains this model locally for a few epochs $E$ on its own dataset $D_i$. After completing local training, selected clients upload their locally trained model weights to the central server, which aggregates them to update the global model as follows:
\begin{equation}
g^{r+1} \leftarrow \frac{1}{K} \sum_{K=1}^{K} g_i^r .
\end{equation}

Fig.~\ref{fig:fl_example} illustrates the architecture of the FL setting in this study, consisting of three main planes: The Data Plane, Programmable Computing Plane, and Aggregation Plane. The Data Plane, representing the IoT environment, gathers data from connected IoT devices for model training. The Programmable Computing Plane includes $K$ edge cloudlets that act as clients, which manage traffic using SDN controllers and locally train the DL. Each client $i$ uses REDUS for reducing the training time, thus reducing $LTT_i$.  At the top, the Aggregation Plane features a central cloud server that utilizes the FedAvg method to aggregate the local models uploaded by the edge cloudlets at each round of training method, updating the global model, which is then redistributed to the cloudlets for further training. 

\begin{figure*}[!t]
    \centering
    \includegraphics[width=\linewidth]{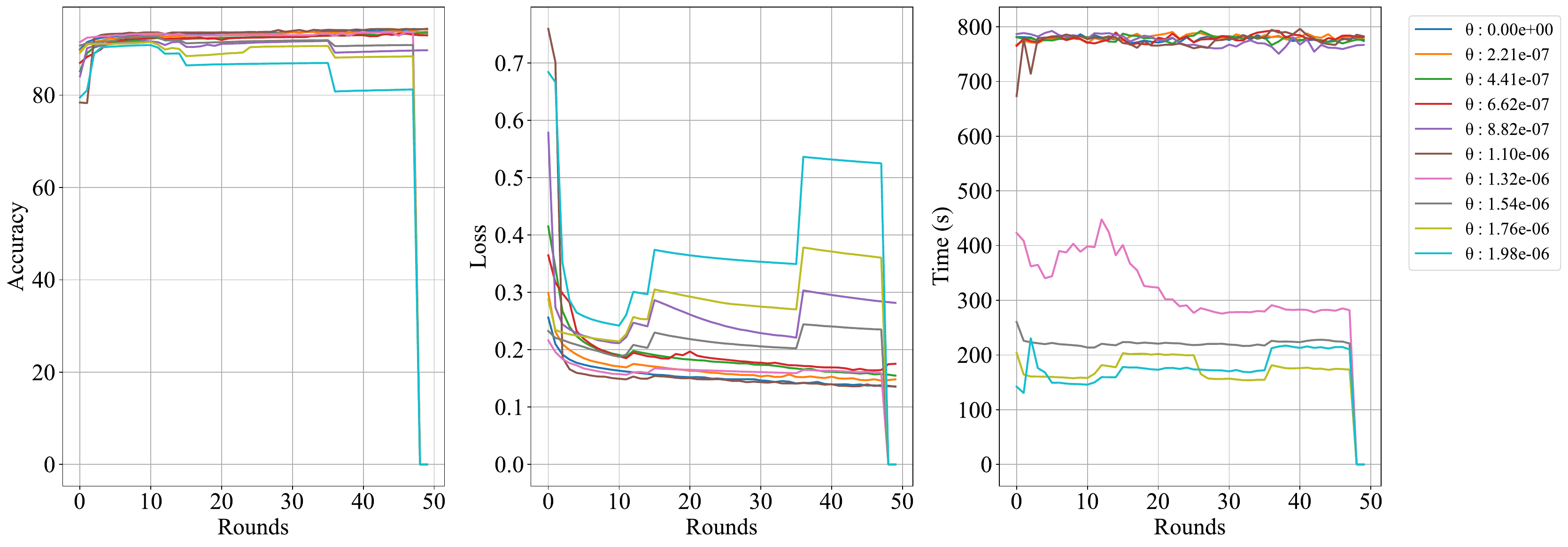}
    \vspace{-0.5cm}
    \caption{The performance metrics achieved on the CICIoT2023 dataset with different threshold values.
\label{fig:ciciot_comparison}}
    \vspace{-0.2cm}
\end{figure*}

\section{Experiments and Results}
\label{sec:exp_res}
We conducted an experiment using the CICIoT2023 dataset for IoT attack detection, repeating it five times and averaging the results. We followed the preprocessing and independent, identically distributed (IID) partitioning as outlined in \cite{eyad24fl}. Each of the five clients held 336K samples ($n$) for training and 105K for testing, while the central server used FedAvg to aggregate local models and evaluated the aggregated model on a test set of 525K samples over 50 communication rounds. Each client trained an Artificial Neural Network (ANN) for 10 epochs using REDUS, with a batch size of 32, an initial learning rate of 0.01, cross-entropy loss, and SGD as the optimizer.

The ANN architecture consisted of five fully connected layers: an input layer flattening 46 features, followed by hidden layers with 256, 512, 256, and 128 neurons, and an output layer with 34 neurons for classification. Each hidden layer used ReLU activation and a dropout rate of 0.2 for regularization.

Threshold values for REDUS were sampled across 10 steps from 0 to $\frac{2}{3} \times \frac{1}{n}$, where a threshold of 0 corresponds to the vanilla ANN model without REDUS. For each threshold, performance metrics of the central server and local training time at each client were averaged across the five experiments, yielding comprehensive time and performance metrics at the round level. For example, the time taken for round 1 at a specific threshold is the sum of the local training times for all five clients at that threshold, averaged across the five experiments.

As for the results, Table \ref{tab:results_table} shows the performance of the CICIoT2023 dataset using the REDUS resampling method with various threshold values ($\theta$) applied during model training. The table is organized to compare each threshold value's performance metrics—accuracy (Acc), loss, and average training time in seconds—with the baseline model (no REDUS applied, $\theta = 0$). The performance metrics include accuracy as a percentage, cross-entropy loss, and average time taken per epoch. Additionally, the table compares each threshold setting with the baseline model by displaying the percentage reductions in both training time and accuracy. The baseline model achieves 94.33\% accuracy, a loss of 0.136, and an average training time of 777 seconds.

Upon analysis, it’s shown that the optimized threshold values $\theta = 1.3 \times 10^{-6}$ and $\theta = 1.5 \times 10^{-6}$ significantly reduce the training time by 59.84\% and 72.60\%, respectively, while maintaining accuracy close to the baseline (93.65\% and 92.71\%, with accuracy reductions of only 0.68\% and 1.62\%). 

\begin{table}[!t]
  \caption{CICIoT 2023 Dataset Performance with REDUS.}
  \label{tab:results_table}
  \centering
  \resizebox{\columnwidth}{!}{\renewcommand{\arraystretch}{1.4}%
    \Large
  \begin{tabular}{l|ccc|cc}
    \toprule
    \multicolumn{1}{c|}{\textbf{REDUS}} & \multicolumn{3}{c|}{\textbf{Performance Metrics}} & \multicolumn{2}{c}{\textbf{Comparison to Baseline}} \\
    \textbf{Threshold $\theta$} & \textbf{Acc (\%)} & \textbf{Loss} & \textbf{Avg. Time (s)} & \textbf{Time Red. (\%)} & \textbf{Acc Red. (\%)} \\
    \midrule
     $0$ (Baseline)       & 94.33 & 0.136 & 777 & N/A & N/A \\
     $2.2 \times 10^{-7}$ & 93.97 & 0.146 & 780 & -0.35 & 0.36 \\
     $4.4 \times 10^{-7}$ & 93.47 & 0.155 & 778 & -0.10 & 0.86 \\
     $6.6 \times 10^{-7}$ & 93.34 & 0.164 & 778 & -0.13 & 0.99 \\
     $8.8 \times 10^{-7}$ & 91.77 & 0.221 & 773 & 0.49 & 2.57 \\
     $1.1 \times 10^{-6}$ & 94.33 & 0.136 & 773 & 0.52 & 0.0 \\
     $1.3 \times 10^{-6}$ & 93.65 & 0.161 & 312 & 59.84 & 0.68 \\
     $1.5 \times 10^{-6}$ & 92.71 & 0.188 & 213 & 72.60 & 1.62 \\
    $1.7 \times 10^{-6}$ & 91.52 & 1.789 & 167 & 78.42 & 2.82 \\
    $1.9 \times 10^{-6}$ & 90.85 & 1.813 & 171 & 77.95 & 3.48 \\
    \bottomrule
  \end{tabular}%
  }
\end{table}

However, higher thresholds (e.g., $1.7 \times 10^{-6}$ and $1.9 \times 10^{-6}$) lead to substantial time savings but a noticeable drop in accuracy, indicating a trade-off between training efficiency and model performance. The optimized thresholds, $\theta = 1.3 \times 10^{-6}$ and $\theta = 1.5 \times 10^{-6}$, are therefore ideal as they maintain a high accuracy level while considerably reducing computational time compared to the baseline. More details and insights can be seen in Fig.~\ref{fig:ciciot_comparison}. 

\section{Conclusion}
\label{sec:conclusion}

This study introduced REDUS, a resampling technique designed to optimize DL training in both centralized and decentralized learning settings. Integrated into FL, REDUS enhances the feasibility of deploying DL models on SDN-managed IoT networks by conserving energy, reducing latency, and sustaining high performance. Experiments on the CICIoT2023 dataset demonstrated that REDUS significantly reduces training time with minimal accuracy loss at optimized thresholds. These results highlight REDUS's potential to improve both the efficiency and scalability of FL in real-world IoT applications, showing that targeted data resampling effectively balances computational demands and model performance. REDUS is especially useful in resource-limited scenarios, such as FL and IoT, where devices usually have constrained computational resources, making it a practical choice for reducing training budgets in these contexts. 

Future work could explore extending REDUS to additional FL security mechanisms. Notably, insights from the coauthor’s thesis on robust and resource-efficient federated learning for IoT security~\cite{Eyad-thesis-2024} provide a foundation for developing more adaptive and scalable frameworks.

\balance
\bibliographystyle{IEEEtran}
\bibliography{refs.bib}

\end{document}